\documentclass[twocolumn,
notitlepage,
aps,
prx,
floatfix,
superscriptaddress
]{revtex4-2}
\usepackage{amsmath, amsfonts, amsthm, amssymb, bbm}
\usepackage[margin=1in]{geometry}
\usepackage{graphicx,xcolor,tikz}
\usepackage[colorlinks=true, hyperindex, breaklinks, linkcolor=blue, urlcolor=blue, citecolor=blue]{hyperref}
\usepackage{physics}

\makeatletter
\newcommand*{\balancecolsandclearpage}{%
  \close@column@grid
  \clearpage
  \twocolumngrid
}
\makeatother

\graphicspath{{fewer_detections/figs/}}

\begin{document}

\preprint{APS/123-QED}

\title{Magic State Injection with Erasure Qubits}

\author{Shoham Jacoby}
\affiliation{AWS Center for Quantum Computing, Pasadena, CA 91125, USA}
\affiliation{Racah Institute of Physics, The Hebrew University of Jerusalem, Jerusalem 91904, Givat Ram, Israel}
\author{Yotam Vaknin}
\affiliation{AWS Center for Quantum Computing, Pasadena, CA 91125, USA}
\affiliation{Racah Institute of Physics, The Hebrew University of Jerusalem, Jerusalem 91904, Givat Ram, Israel}
\author{Alex Retzker}
\affiliation{AWS Center for Quantum Computing, Pasadena, CA 91125, USA}
\affiliation{Racah Institute of Physics, The Hebrew University of Jerusalem, Jerusalem 91904, Givat Ram, Israel}
 \author{Arne L. Grimsmo}
\affiliation{AWS Center for Quantum Computing, Pasadena, CA 91125, USA}

\date{\today}

\begin{abstract}
    Erasure qubits constitute a promising approach for tackling the daunting resources required for fault-tolerant quantum computing. By heralding erasure errors, both the error-correction threshold and the sub-threshold scaling of the logical error rate are significantly improved.
    While previous research has focused primarily on fault-tolerant quantum memories,
    we extend this investigation to magic state injection—a critical yet resource-intensive component of fault-tolerant quantum computation.
    We show that, after postselecting on erasures, the logical error rate of the injected magic state is set by the residual Pauli error,
    while the space-time overhead is only marginally increased as compared to non-erasure qubits with a similar noise strength.
    These conclusions hold both for injection into the surface code, and for injection and cultivation on the color code.
    For the former, we show that most of the gains can be achieved by using just three strategically placed erasure qubits in the surface code patch, independent of the patch size. For the latter, in contrast, it is beneficial to have all the qubits in the cultivation patch be erasure qubits.
    Our results for cultivation suggest that algorithmically relevant logical error rates may be within reach \emph{without} magic state distillation for erasure rates $\lesssim 4\times 10^{-3}$ and residual Pauli error rates $\sim 10^{-4}$. 
\end{abstract}
\maketitle
\section{Introduction}

One of the foremost challenges in realizing universal quantum computation is the reliable implementation of non-Clifford operations, such as the $T$ gate. Magic state based approaches represent a leading strategy for enabling these crucial gates in fault-tolerant architectures
\cite{bravyi2012magic}. However, magic states can often not be created fault-tolerantly and instead must be injected into the code and distilled to higher fidelity \cite{bravyi2012magic, Horsman_201_fowler_lattice_injection,Fowler_2012, litinski2019magic}.

The fidelity with which these states are created has a large impact on the resource cost of the subsequent distillation process.
For example, in the 15-to-1 distillation process \cite{bravyi2012magic,litinski2019magic}, an injection error rate $p$ leads to an output error rate $\sim p^3$ after distillation.
This means that an order-of-magnitude improvement in the injection infidelity results in a three-order-of-magnitude improvement in distillation fidelity. 
In particular, the difference in overhead for magic state injection with $p\sim 10^{-3}$ vs. $p\sim 10^{-4}$ error rates can be striking, because it may determine whether one or multiple distillation rounds are necessary for near-term fault-tolerant applications~\cite{litinski2019magic}. Similar conclusions hold for the recently introduced ``Magic State Cultivation'' protocol, where $\sim 10^{-4}$ physical error rates bring near-term applications within reach \emph{without} distillation~\cite{gidney2024magic}.

Magic state injection protocols have been improved over a series of works~\cite{Horsman_201_fowler_lattice_injection,li2015magic, lao2022magic, gidney_hook_injection, Singh_2022XZZX_inject,chamberland2020very, bombin2024fault, gidney2024magic}. Nevertheless, injection is fundamentally limited by the error rate of the underlying physical qubits.
This limitation motivates the search for new qubit platforms that naturally support higher-fidelity operations, or, as is the case for erasure qubits, offer mechanisms to detect errors more efficiently.

Erasure qubits have emerged as a promising candidate for the implementation of fault-tolerant quantum computation.
In erasure-based architectures, qubits can signal whether an error has occurred, effectively converting unknown errors into heralded erasure errors.
This additional information enables more targeted error-correction strategies, leading to higher thresholds for fault-tolerant memories~\cite{Wu_2022atoms, Kubica_2023}.
Potential realizations of erasure qubits include neutral atoms \cite{Wu_2022atoms}, trapped ions \cite{Kang_2023ions} and superconducting qubits \cite{Kubica_2023,Teoh_2023cavities}, with recent experimental demonstrations
\cite{Ma_2023atom_exp,Scholl_2023atom_exp,koottandavida2024erasure,Levine_2024}.

In this work, we extend magic state injection circuits on the surface \cite{bravyi1998quantum, dennis2002topological,kitaev2003fault,li2015magic, lao2022magic, gidney_hook_injection} and color code \cite{chamberland2020very, gidney2024magic} to erasure qubit implementations. Many of our conclusions are of a general nature and should also apply to other injection protocols.
When post-selecting on erasure events, the logical injection error is independent of the erasure error rate, and is set by the (assumed much lower) residual Pauli error rate within the computational subspace of the erasure qubits. We show that this conclusion continues to hold to a good approximation for noisy erasure detection at realistic error rates.

Moreover, there is only a marginal reduction in the injection acceptance rate, if we compare erasure qubits to non-erasure qubits and set the erasure rate of the former equal to the Pauli error rate of the latter.
In other words, the conversion of Pauli errors to heralded erasure errors comes at minimal cost in terms of the number of retries before a magic state is accepted.
A second benefit comes from the well-known improved error-correction performance associated with erasure qubits \cite{Wu_2022atoms, Kubica_2023}, which reduces the cost of expanding to a larger code after the magic state has been injected, making it practical to perform injection on smaller distance codes.

In contrast to quantum error correction, where frequent erasure checks and resets are required to gain a substantial advantage over non-erasure qubits~\cite{Bailey_optimization_2024}, the post-selected nature of state injection means we can check for erasures only at the end of the injection. Alternatively, a practical choice which we use in our numerics, is to check once per syndrome round during the injection, in parallel with ancilla measurements.
Less frequent erasure checks may relax hardware requirements, as erasure detection and reset can be slow in practice.
The challenges associated with rapid erasure detection can perhaps be avoided, while still retaining very large overhead reduction for magic state factories, by using specialized patches with erasure qubits for injection, with magic states subsequently transported to patches of non-erasure qubits.

Interestingly, we show that almost all the benefit of using erasure qubits in the injection step is retained by placing only a small, constant number of erasure qubits at strategic locations in the code patch. For surface code injection, as few as three erasure qubits are required, independent of the desired size of the final encoded magic state. For the ``cultivation'' step following the initial injection in 
 the recently introduced ``Magic State Cultivation'' protocol~\cite{gidney2024magic}, the situation is a bit different. There, it does pay to make most of the qubits in the patch into erasure qubits, as we show in Sec.~\ref{sec:Cultivation}.

We demonstrate the advantages of erasure qubits through numerical simulations using the injection protocol presented in Refs.~\cite{lao2022magic} and~\cite{gidney_hook_injection} on the surface code, as well as the recently introduced injection and cultivation protocol on the color code~\cite{gidney2024magic}. We compare erasure qubits with varying erasure rates to non-erasure qubits. 
The results reveal that even at relatively high erasure rates and noisy erasure detection, there is a significant advantage for injection using erasure qubits.

\section{Erasure Qubits}
Erasure qubits are a type of qubit that converts a specific noise mechanism to a heralded error. Erasure can be detected by measuring the qubit in a state outside of its computational subspace \cite{Kubica_2023, Levine_2024, Wu_2022atoms,koottandavida2024erasure}, or by using an ancilla that flags erasure events while the data qubits remain in the computational subspace~\cite{Teoh_2023cavities}. 

Erasure qubits become beneficial when erasures are the dominant error mechanism in the system and undetectable errors are heavily suppressed. In the limit where erasure dominate, the surface code error-correcting threshold is close to $5\%$ erasure per CNOT gate \cite{Wu_2022atoms,Kubica_2023, Bailey_optimization_2024}, and the sub-threshold logical error rate has better scaling with distance~\cite{Wu_2022atoms, Bailey_optimization_2024}.
More concretely, for small physical error rates we expect the logical error to be exponentially suppressed with distance as~\cite{Fowler_2013,Bravyi_2013,Watson_2014,Bailey_optimization_2024}
\begin{equation} \label{eq:ansatz-erasure}
    p_{L} \propto q^{\alpha d},
\end{equation}
where $q$ approaches a linear function of the physical error rates in the low noise regime ($q \ll 1$) and $q=1$ marks the threshold. While for non-erasure qubits we have $\alpha=1/2$, the scaling is generally improved for erasure qubits, and approaches $\alpha=1$ in the limit where erasures dominate~\cite{Bailey_optimization_2024}.

Multiple experiments have demonstrated that one can engineer a significant bias where undetectable errors within the computational subspace are far less likely than erasure errors. For superconducting transmon qubits, the recently introduced dual-rail qubit~\cite{Kubica_2023,Teoh_2023cavities} mostly suffer from energy-relaxation noise, while dephasing is suppressed by a stable energy gap due to a capacitive coupling of the two transmons comprising the dual-rail. With superconducting cavity dual-rails, it is possible to detect noise on the transmon ancilla, which is the central source of noise in the system. Lastly, for cold atoms, decay from the Rybderg state is significantly biased toward the identifiable meta-stable states, meaning over $98 \%$ of the decay during a two-qubit gate can be detected \cite{Wu_2022atoms}.

\section{\label{sec:injection}Magic State Injection}

\begin{figure}
    \centering
    \includegraphics[width=\linewidth]{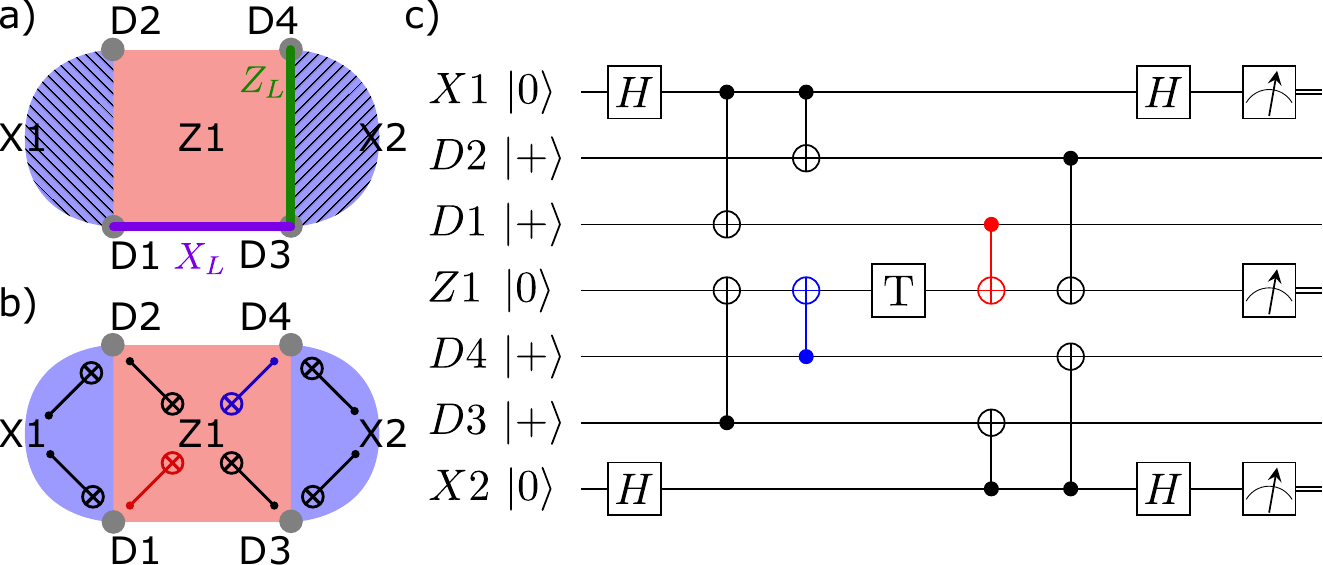}
    \caption{The first step of the hook injection circuit from Ref.~\cite{gidney_hook_injection} for a rotated surface code of distance $d_1=2$. a) physical layout of the four data qubits D1-D4, the two $X$ stabilizers (blue) and the Z stabilizer (pink). The hatched stabilizers are deterministic in the first round. The logical operators $Z_L$ and $X_L$ are depicted in green and purple, respectively. b) CNOT locations over the surface code patch. c) The circuit implementing the first round of the hook injection. Some physical Pauli errors on the $T$ gate and blue and red CNOTs can flip the logical operator without being detected, and thus their error probability has a linear contribution to the logical error rate. The $T$ gate translates to a rotation of the $Z_L$ logical operator leading to the preparation of a $|T\rangle=T|+\rangle$ state (Appendix \ref{sec:hook-inject-appendix}).}
    \label{fig:hook-injection}
\end{figure}

The Eastin-Knill theorem \cite{eastin2009restrictions} states that universal quantum computation cannot be implemented transversely on a quantum error-correcting (QEC) code. To overcome this limitation, researchers have developed various schemes, including code-switching and gate teleportation.

Gate teleportation requires specific input states that typically cannot be prepared fault-tolerantly. In practice, this challenge is addressed through a two-step process: (1) Magic state injection \cite{horsman2012surface,li2015magic, lao2022magic,gidney_hook_injection}, where many noisy encoded states are generated in a non-fault-tolerant manner, and (2) magic state distillation \cite{Bravyi_2005}, where noisy states are distilled into fewer high-fidelity states suitable for gate teleportation.

Magic state injection involves encoding a magic state in a low-distance ($d_1$) code and subsequently expanding it to a code with a larger distance ($d_2$). This method is particularly vulnerable to errors during the initial stages of the process. These early errors can propagate through the system and persist, regardless of the final code size.

Here, we consider the surface code magic state injection protocol described in \cite{gidney_hook_injection, lao2022magic, li2015magic}, but we emphasize that many of our conclusions are general and should apply to other injection schemes as well. Color code cultivation is presented in Sec.~\ref{sec:Cultivation}. The protocol proceeds as follows:

1. A noisy magic state is prepared in a $d_1$ surface code. We exemplify this step by implementing both the scheme from Lao and Criger in Ref.~\cite{lao2022magic}, building on previous work by Li~\cite{li2015magic}, which we refer to as ``Lao-Criger injection,'' and the scheme from Gidney in Ref.~\cite{gidney_hook_injection}, which we refer to as ``hook injection.''
Lao-Criger injection allows for injection of an arbitrary state, while hook injection allows for the injection of states confined to the $XY$ plane ($\bra{\psi}Z\ket{\psi}=0$). 
This step includes initialization of the data qubits and performing one round of stabilizer extraction. A subset of stabilizers are deterministic in the absence of errors, and the process is restarted if any of these give an unexpected outcome.

2. Stabilizer measurements are performed on the distance $d_1$ surface code for another $r-1$ rounds. If any stabilizer measurement flips in subsequent rounds, the state is discarded and the process restarts. 
In earlier work $r=2$ was typically used~\cite{lao2022magic,li2015magic}, but in Ref.~\cite{gidney_hook_injection}, it was shown that for large values of $d_1$ increasing $r$ is beneficial.

3. The surface code is finally expanded to a larger distance $d_2$,
followed by $d_2$ rounds of standard error correction.
The final distance, $d_2$, is chosen to be large enough to support the desired error rate of the subsequent distillation protocol \cite{litinski2019magic}. 

The logical error rate has a contribution from undetected low-weight errors in the first two steps, and topologically non-trivial errors in the third step.
In Fig. \ref{fig:hook-injection} we illustrate the first step of the hook injection protocol for the case of a $d_1=2$ surface code, and we present a full analysis of the two injection strategies in Appendix \ref{sec:injection-types}.

During the first step, single fault locations can cause an undetected logical error (see, e.g., the $T$ gate and the blue- and red-colored CNOTs in~Fig.~\ref{fig:hook-injection})
\footnote{This can be understood by considering the initialization of the magic state as some physical pulse with specific time and amplitude, that can in principle rotate to any arbitrary angle. Any error in the rotation angle would result in some overlap of the state on a non-magic state that can not be identified without distillation.}.
Consequently, the logical error rate has a linear contribution from the Pauli error probability of one-qubit (1Q) gates $p_1$, two-qubit (2Q) gates $p_2$ and initialization (IN) $p_\text{IN}$.

To quantify these contributions, we approximate the logical error rate achieved after the first two steps as follows:
\begin{equation} \label{eq:logical-error-rate}
    p_{L,\text{inj}} \qty(\bar{p}) \approx  a p_1 + b p_2 + c p_\text{IN} + O(p^2) \;.
\end{equation}
with $\bar{p} \equiv \qty(p_1,p_2,p_\text{IN})$.
Here, $a, b, c \in \mathbb{R}$ are coefficients derived from counting the number of undetected fault locations that flips the logical operators. Coefficients for hook injection and Lao-Criger injection are given in Appendix \ref{sec:injection-types}. For a noise model with $p_\text{IN}=p_2=p$, $p_1=p/10$ and injection of the $\ket{S}=(\ket{0} + i\ket{1})/\sqrt 2$ state [which can be simulated by Clifford operations, unlike the magic state $|T\rangle = (\ket{0}+e^{i\pi/4}\ket{1})\sqrt 2$] the logical error rate is $p_\text{HI} = 7/30p$ for the hook injection and $p_\text{LI}=46/30 p$ for Lao-Criger injection, showing the favorable performance of hook injection.

\subsection{Injection with Erasure Qubits} \label{sec:injection-with-erasure}
By implementing magic state injection using erasure qubits \cite{Kubica_2023}, we can use the additional information received from the erasures to improve the post-selection.

\begin{figure}
    \centering
    \includegraphics[width=\linewidth]{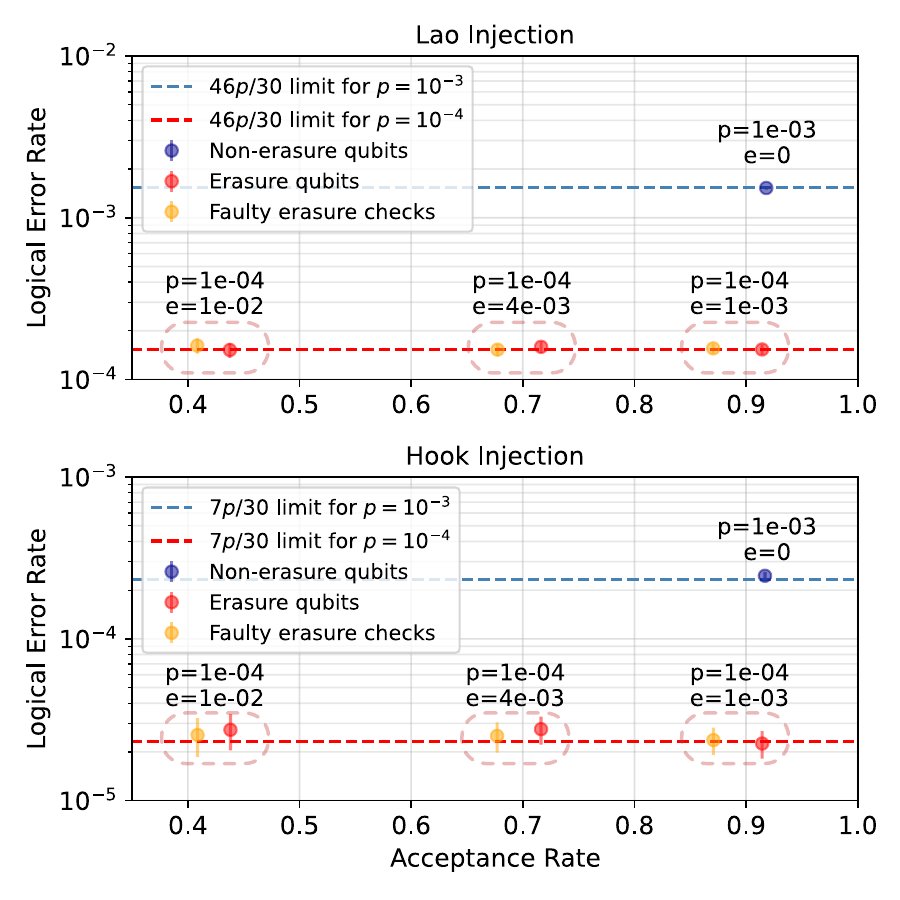}
    \caption{Logical error rate and acceptance rate for ungrown injection with $d_1=3$ for Lao-Criger injection and hook injection of a $|S\rangle = S|+\rangle$ state.
    Points represent non-erasure ($e=0, p_n=10^{-3}$), almost perfect erasure conversion ($e=10^{-3}, p=10^{-4}$), and two values of non-perfect erasure conversion rates ($e=4 \cdot 10^{-3}, p=10^{-4}$ and $e=10^{-2}, p=10^{-4}$).
    The theoretical limits [Eq. \eqref{eq:logical-error-rate}] for non-erasure qubits and erasure qubits are depicted by the dashed blue and red lines, respectively.
    Each erasure scenario is calculated for perfect erasure detection and for erasure detection error rate of $e_\text{FP}=e_\text{FN}=10^{-2}$ for false-positive and false-negative detections, respectively.
    }
    \label{fig:ungrown-injection-rates}
\end{figure}

Erasure qubits split the noise into two parts: undetectable Pauli errors that contribute to the injection fidelity in linear order, and erasure errors that are identifiable using erasure detection. 
Useful erasure qubits should convert dominant noise processes to erasures. We therefore assume that the probabilities $e_1$, $e_2$, and $e_\text{IN}$, of erasure errors during 1Q, 2Q, and initialization, respectively, obey
\begin{equation} \label{eq: e>>p}
e_2 ,e_1, e_\text{IN} \gg p_2 ,p_1, p_\text{IN},
\end{equation}
for erasure qubits.
 
This allows us to reduce the logical error by discarding the state for every detected erasure in the first two steps, at the expense of lowering the injection acceptance rate.
For perfect erasure detection, such that every erasure in the first round is identifiable, the logical error rate of the injection is strictly given only by the undetectable errors, i.e.,
\begin{equation} \label{eq:erasure-logical-rate}
    p_{L,\text{inj}} \qty(\bar{e}, \bar{p}) =  p_{L,\text{inj}} \qty(\bar{p}) \;.
\end{equation}
independent of $\bar e$ and with $p_{L,\text{inj}}(\bar p)$ defined as before, c.f. Eq.~\eqref{eq:logical-error-rate}.  
This manifests a key advantage of utilizing erasure qubits for magic state injection, as the logical error rate does not depend on the erasure rate. 

On the other hand, discarding states due to erasure errors directly affects the acceptance rate. However, the impact is not very large, if we compare to non-erasure qubits with a similar noise rate, as we show in the following.

The acceptance rate of the protocol can be broken down into two parts: the probability that no erasure has occurred $\text{AR}_e(\bar{e})$, and the probability that no Pauli error was detected $\text{AR}_p(\bar{p})$. 
The overall acceptance rate of an injection circuit discarding erasure errors is then the product
\begin{align}
    \text{AR} \qty(\bar{e},\bar{p}) = \text{AR}_e \qty(\bar{e})
    \text{AR}_p \qty(\bar{p})
    \;.
\end{align}

It is natural to compare the performance of a protocol with erasure qubits to that of using non-erasure qubits, where the Pauli error rate of the non-erasure qubits is comparable in magnitude to the erasure rate. This is motivated by the observation that physical implementations of erasure qubits ``convert'' the dominant physical error processes, that would normally lead to undetectable errors, into detectable erasures~\cite{Ma_2023atom_exp,Scholl_2023atom_exp,koottandavida2024erasure,Levine_2024}.

For clarity, we denote the Pauli error rates when using non-erasure qubits by $\bar p_n$.
Numerical analysis suggests that most mechanisms contributing to Pauli errors are detected (Appendix \ref{sec: success-rate}), and as such we get that for $\bar{p}_n=\bar{e}$ the acceptance rates due to erasure errors and due to Pauli errors are nearly identical, i.e.,
$\text{AR}_e (\bar{e}) \approx \text{AR}_p(\bar{p}_n)$. 
In addition, in the limit of interest where the undetected Pauli error rate of the erasure qubits is very small [Eq.~\eqref{eq: e>>p}], only a small fraction of the discarded states originates from residual Pauli errors, i.e.,
$1-\text{AR}_p(\bar{p}) \ll 1 - \text{AR}_e(\bar{e})$. Consequently, the acceptance rate of erasure qubits is expected to be nearly the same as that of non-erasure qubits when $\bar e = \bar p_n$:
\begin{align} \label{eq:erasure-success-rate}
    \text{AR}(\bar{e}, \bar{p}) \simeq \text{AR}_p(\bar{p}_n=\bar e)
    \;.
\end{align}

In practice, implementing erasure qubits may come at the cost of increased complexity compared to non-erasure qubits in the same physical platform, making an apples to apples comparison challenging.
In particular, erasure qubits may come at the cost of additional hardware, more frequent measurements and resets, and slower gates. 

In this work, we consider three scenarios for comparison: an almost perfect ``conversion'' of Pauli to erasure errors $\bar e=\bar p_n$, and two values for non-perfect ``conversion'' $\bar e=4\bar p_n$ and $\bar e=10\bar p_n$. We note that certain architectures, such as tunable transmons and cold atoms, may be able to achieve $\bar e\approx \bar p_n$~\cite{Wu_2022atoms, Kubica_2023}, depending on the gate implementation.
Additionally, we assume that erasure qubits retain some residual Pauli error $p$, which we take to be an order of magnitude smaller than for non-erasure qubits in our numerics, $p=p_n/10$. 

Fig.~\ref{fig:ungrown-injection-rates} presents a comparison of injection acceptance rates and logical error rates for non-erasure and erasure qubits, at the different ratios of $\bar p_n$ to $\bar e$.
We use a noise model where reset, measurement, and CNOTs carry a probability 
$p_n$ ($p$)
of inducing a Pauli error on the participating non-erasure (erasure) qubits, while other single-qubit gates and idling  carry a reduced error probability of
$p_n/10$ ($p/10$).
For the erasure qubits, we assume equal erasure probability for reset, measurement and CNOT operations 
$e$,
while other single-qubit operations have a reduced erasure probability
$e/10$.
Idling occurs at every stabilizer extraction round and affects all qubits not currently engaged in measurement. 
Erasure detection is performed at the end of every stabilizer extraction round in parallel with ancilla measurements. 
An erased qubit that interacts with another qubit in a two-qubit gate induces a fully depolarizing channel on the other qubit. See Appendix \ref{sec:noise-model} for further details of the noise model.

From Fig.~\ref{fig:ungrown-injection-rates}, we see that for the $e=p_n$ scenario, the injection acceptance rate with erasure qubits is reduced only marginally compared to non-erasure qubits, as we anticipated in Eq.~\eqref{eq:erasure-success-rate}. However, the logical error rate improves significantly. As expected the logical rate only depends on the residual error rate $p$ when erasure detection is perfect [Eq. \eqref{eq:erasure-logical-rate}].
This observation is essentially that post-selection on an event results in an error rate that is independent of the probability of that event~\cite{bombin2024fault,knill2004fault}.

\begin{figure*}
    \centering
    \includegraphics[width=\linewidth]{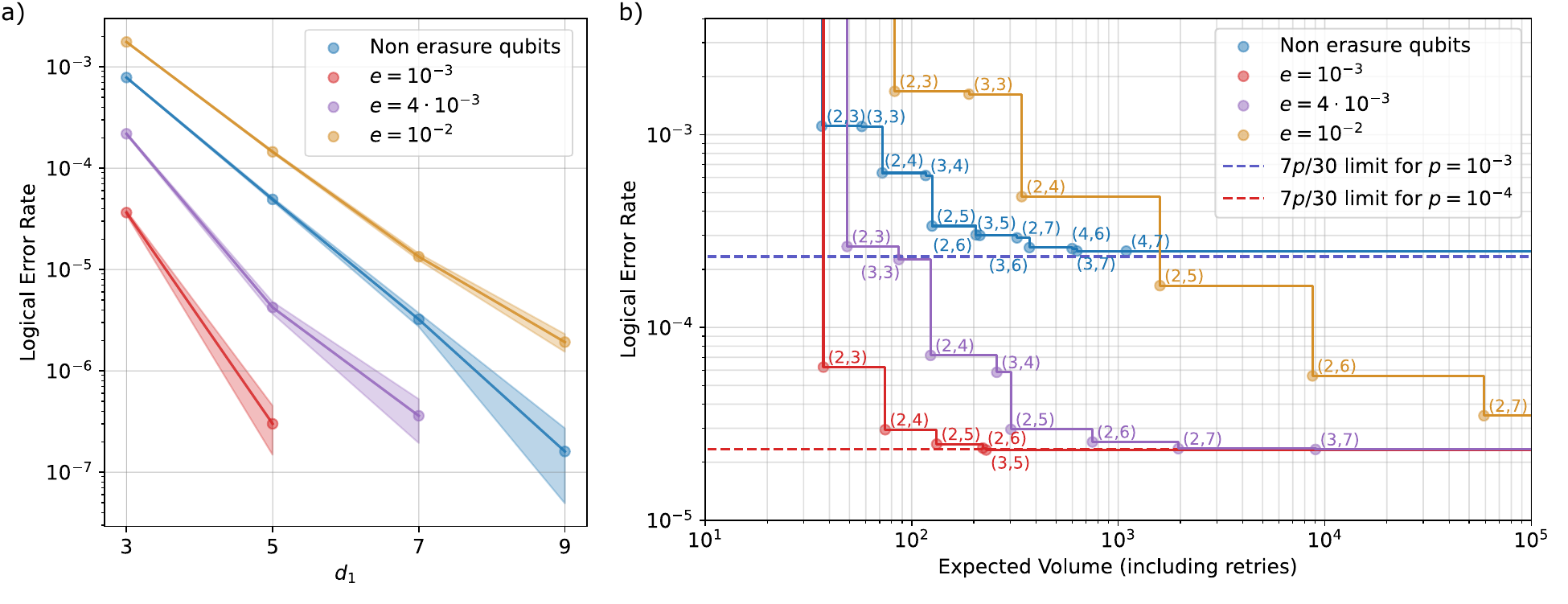}
    \caption{a) Logical error rate of the expansion from a distance $d_1$ to a $d_2=15$ surface code. The encoded $d_1$ surface code state is prepared noiselessly, allowing for examination of the error contributions from the expansion step.
    The same erasure conversion scenarios as in Fig. \ref{fig:ungrown-injection-rates} are analyzed: non-erasure qubits ($e=0, p_n=10^{-3}$), near-perfect erasure conversion ($e=10^{-3}, p=10^{-4}$), and two values non-perfect erasure conversions ($e=4 \cdot 10^{-3}, p=10^{-4}$ and $e=10^{-2}, p=10^{-4}$).
    b) Pareto front of the entire injection+expansion protocol for the same scenarios as in (a) (same color key). The annotation next to each point represents the values of $(r,d_1)$.
    The theoretical limits [Eq. \eqref{eq:logical-error-rate}] for non-erasure qubits and erasure qubits are depicted by the dashed blue and red lines, respectively.
    }
    \label{fig:expansion-and-end2end}
\end{figure*}

\subsubsection{Faulty erasure detection}

For imperfect erasure detection, false positive and false negative detections contribute differently.
False positive erasure detections only lead to unnecessary discards of potentially valid states, reducing the protocol's acceptance rate. 
On the other hand, false negative detections can contribute to the logical error rate. However, the latter is mitigated by three factors: first, this is a second order effect where an erasure first happens and then goes undetected, such that the probability of false negatives is already quite low; second, these undetected erasure errors generate depolarizing noise on both themselves and interacting qubits in our model, making them likely to trigger a stabilizer detection event; third, they must evade detection across all $r$ rounds.

To balance the competing effects between false positive and false negative erasure detections, we choose to discard the state only if we detect two erasures. This approach reduces unnecessary discards from false positives, as a single erroneous detection will not trigger abortion. While this could potentially allow some true erasure errors to go undetected (false negatives), we find that this is unlikely in practice, as undetected erasure errors are likely to trigger stabilizer detection events as they stay erased and inject noise into the system over multiple rounds.
We have found in our numerical analysis that this trade-off is favorable, as the reduction in false-positive-induced discards outweighs the minimal increase in undetected errors.

As demonstrated in Figure \ref{fig:ungrown-injection-rates}, introducing significant false negative and false positive rates $e_\text{FN}=e_\text{FP}=10^{-2}$ with this scheme results in a minimal effect on the logical error rate, and only a marginal reduction of the acceptance rate. Eqs.~\eqref{eq:erasure-logical-rate} and~\eqref{eq:erasure-success-rate} are thus reasonably good approximations even for noisy erasure detection.

\subsection{Code expansion} \label{sec: expanding-the-state}

In the third step we expand the code from distance $d_1$ to distance $d_2$, where $d_2$ is the desired final distance of the surface code.
The overall error rate of the injection protocol has contributions from errors in all three steps, 
which can be approximately expressed as
\begin{align}
    p_L \simeq p_{L,\text{inj}} + p_{L,\text{exp}}\;,
\end{align}
with $p_{L,\text{inj}}$ and $p_{L,\text{exp}}$ denoting the logical error rates in the injection and expansion steps, respectively.

The logical error rate of the injection process is limited by the linear contributions from low weight errors in the initial step [Eq. \eqref{eq:logical-error-rate}].
The expansion step should ideally exhibit a significantly lower error rate compared to the error rate of the injection process.

During the first stabilizer extraction round in the expansion process, the code still exhibits fault distance of the smaller distance $d_1$ code, and we thus expect the logical error $p_{L,\text{exp}}$ to be dominated by errors happening in this round. For low physical error rates, we thus expect $p_{L,\text{exp}}$ to follow Eq.~\eqref{eq:ansatz-erasure} with $d=d_1$ and where $q$ is a linear combination of $e$ and $p$ (see Appendix~\ref{sec:simulating-erasure-qubits} for numerical fits to this ansatz).

By simulating an expansion of a perfectly prepared state (i.e., one with no noise prior to the expansion round), we can isolate and quantify the logical error rate contributed solely by the expansion step, $p_{L,\text{exp}}$.
Figure~\ref{fig:expansion-and-end2end}a illustrates these simulation results across various parameter regimes, demonstrating how expansion-induced errors scale with code distance $d_1$ and physical error rates, with fixed $d_2=15$.
To compare erasure and non-erasure qubits, we use the same noise model as in~\ref{sec:injection-with-erasure}.

\label{subsection:success_rate_simulation}

To evaluate the effectiveness of the complete injection process, we simulate the entire protocol as described in Ref.~\cite{gidney_hook_injection}. The simulation begins with hook injection of an $|S\rangle$ state into a surface with code distance $d_1$, followed by $r-1$ round of stabilizer extraction, a single round of expansion to distance $d_2$, $d_2$ memory rounds, and ends with in-place $Y$ basis measurement \cite{Gidney_2024inpace-Y} to determine the injection success. In this section, we assume perfect erasure detection after every gate, in order to simplify the decoding process. We refer to \cite{Bailey_optimization_2024} for detailed analysis of how sparse and non-perfect erasure detection can affect the results of a quantum memory protocol.

In Fig. \ref{fig:expansion-and-end2end}b, we plot the error rate and the expected volume of the entire injection circuit, including expansion.
The expected volume is calculated as:
\begin{align}
    V \qty(d_1, r)
    =
    \frac{Q_{d1} r}{\text{AR}}
    \;,
\end{align}
with $Q_{d_1}$ being the number of data and ancilla qubits in the distance $d_1$ surface code and $\text{AR}$ being the acceptance rate of the injection. The error rate has contributions from all of the injection steps, and as such, we can see that for lower $d_1$ values, the expansion is limited by the errors in the expansion round. For higher values of $d_1$, the logical rate approaches the ideal logical error rate of $7/30p$ [Eq. \eqref{eq:logical-error-rate}].

When comparing the logical error rates shown in Fig. \ref{fig:expansion-and-end2end}a for the expansion step alone and Fig. \ref{fig:expansion-and-end2end}b for the full injection process, it becomes evident that the limiting factor depends on which stage contributes more significantly to the total error. If the dominant contribution arises from the expansion step because $d_1$ is not sufficiently large, then no matter how much the injection stage is improved, the final logical error rate remains constrained by the expansion process itself. On the other hand, if the injection step is the primary source of errors, the overall logical error rate quickly saturates at the theoretical limit of $7/30p$ [Eq. \eqref{eq:logical-error-rate} and Appendix~\ref{sec:injection-types}].

\section{Hybrid erasure and non-erasure system} \label{sec:Hybrid-architecture}

\begin{figure}
    \raggedright
    \includegraphics[width=\linewidth]{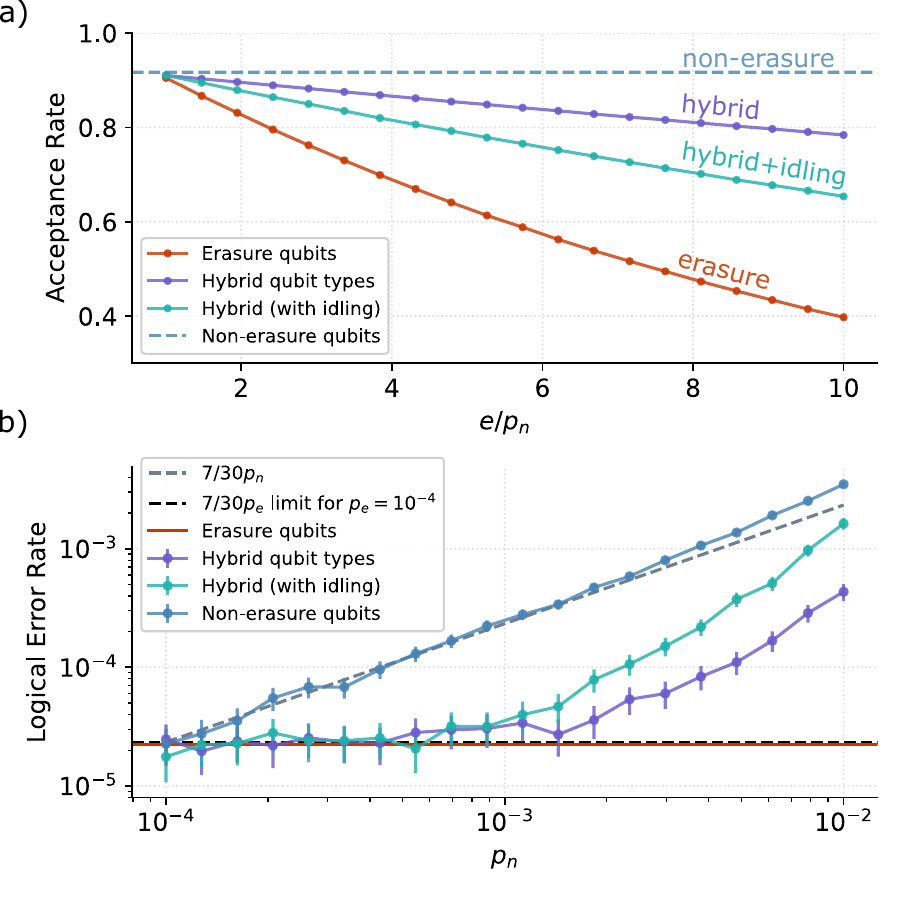}
    \caption{
    (a) Acceptance rates and (b) logical error rates for ungrown qubit patches using different architectures: all erasure qubits (red), all non-erasure qubits (blue), a hybrid design combining erasure and non-erasure qubits (purple), and a hybrid design while taking into account different gate times (green). The results demonstrate that for practical non-erasure error rates ($p_n \sim 10^{-3}$), hybrid architectures maintain low logical error rates comparable to the erasure-only case, while achieving higher acceptance rates. $p_e = 10^{-4}$ is constant throughout.
    }
    \label{fig:hybrid-patch-results}
\end{figure}

In certain hardware implementations, such as superconducting qubits, realizing erasure-enabled qubits can be more expensive than non-erasure qubits~\cite{Levine_2024}. 
Motivated by this, we investigate a hybrid strategy that integrates both erasure and non-erasure qubits within the same surface code patch, deploying erasure qubits only at those circuit locations where they provide the most benefit.

The injection circuit is subject to two distinct types of fault locations: undetectable faults that add directly to the logical error rate and detectable faults that cause the state to be discarded and the injection procedure to be restarted, thereby lowering the overall acceptance rate without affecting the logical error rate.
By introducing erasure qubits to circuit locations that contribute to the first (undetectable) error type, and using non-erasure qubits elsewhere, one can retain the reduced logical error rate of erasure qubit injection while minimizing the number of required erasure qubits.
Furthermore, in scenarios in which erasure qubits exhibit higher error rate compared to non-erasure qubits, $e>p_n$, the reduced error rate associated with the second (detectable) type of error mechanism enhances the protocol’s acceptance rate by reducing the frequency of discards.
In this section, we add a subscript to the residual Pauli error rates of the erasure qubits, $p_e$, to better distinguish those from the Pauli error rate of the non-erasure qubits in the same code patch, which we denote by $p_n$.

The resulting error rate injection on the hybrid patch preserves the linear contributions from $p_e$, as it is driven by the residual Pauli error of the erasure qubits participating in the ``sensitive" circuit locations. Additionally, there is an extra contribution proportional to $p_n^2$, which is negligible as long as $p_n^2 \ll p_e$:
\begin{align}
    p_{L,\text{inj},\text{hybrid}}(p_e,p_n) = p_{L,\text{inj}} (p_e) + O(p_n^2)
    \;.
\end{align}

For example, in the hook injection circuit depicted in Figure \ref{fig:hook-injection}, our proposed approach is implemented by declaring $\text{D1}$, $\text{D3}$ and $\text{Z1}$ as erasure qubits, as they participate in the colored CNOT gates and in the $T$ gate. These three qubits covers all fault locations that contribute linearly to the logical error rate, independent of the size of the surface code patch.

In some architectures, and specifically superconducting qubits, erasure qubits exhibit longer gate times, requiring the non-erasure qubits to remain idle during this period.
We mimic such a scenario by comparing to an alternative noise model, inspired by the relaxation times of the superconducting dual-rail qubits and transmons (see Appendix~\ref{sec:Hybrid-dual-rail-and-transmon-architecture}). 
This alternative noise model has a CNOT Pauli error rate for non-erasure qubits given by
$p_2^* = \qty(p_{n} + e_2)/2$.
This alternative noise model is labeled ``Hybrid (with idling)'' in Fig.~\ref{fig:hybrid-patch-results}.

We compare numerical results for hook injection with non-erasure qubits, erasure qubits, and the hybrid approach---where only three qubits are erasure qubits---in Fig.~\ref{fig:hybrid-patch-results}.
In Fig.~\ref{fig:hybrid-patch-results}a, we show the acceptance rate for the different scenarios.
In Figure \ref{fig:hybrid-patch-results}b, we show the corresponding logical error rate as function of the erasure conversion ratio $e/p_n$. 
It can be seen that the hybrid architecture shows a significant acceptance rate improvement over all-erasure patch, while the logical error rate is essentially unaffected as long as $p_n^2 \ll p_e$, as expected. This could thus be a particularly cost-effective way to reduce magic state overhead in architectures where high quality erasure qubits can be realized, but are more expensive than non-erasure qubits.

\section{Injection and Cultivation on the color code} \label{sec:Cultivation}

In a recent paper, Gidney, Shutty, and Jones, introduced a novel approach for creating magic states, dubbed ``Magic State Cultivation'' \cite{gidney2024magic}. This procedure comprises three steps: injection, cultivation, and grafting, where the latter is a code expansion step from a small color code to a hybrid color and surface code. Compared to the injection and expansion protocols discussed in the preceding two sections, the cultivation step is a fundamentally new ingredient.

The injection step prepares a magic state in a small color code, similarly to the surface code injection protocols discussed in Sec.~\ref{sec:injection}. We here focus on the ``unitary injection'' from Ref.~\cite{gidney2024magic}, which shares similarities with hook injection for the surface code, consisting of a state creation phase via unitary operation followed by a round of stabilizer measurement.
This step results in a magic state that has a linear physical error rate contribution to the logical error rate:
\begin{align}\label{eq:logical-error-rate-color}
    p_{L,\text{color},\text{inj}} = \frac{2}{15}p_2 + \frac{1}{3}p_1 \;,
\end{align}
using the same notation for one-qubit and two-qubit locations as previously.

\begin{figure}
    \centering
    \includegraphics[width=\linewidth]{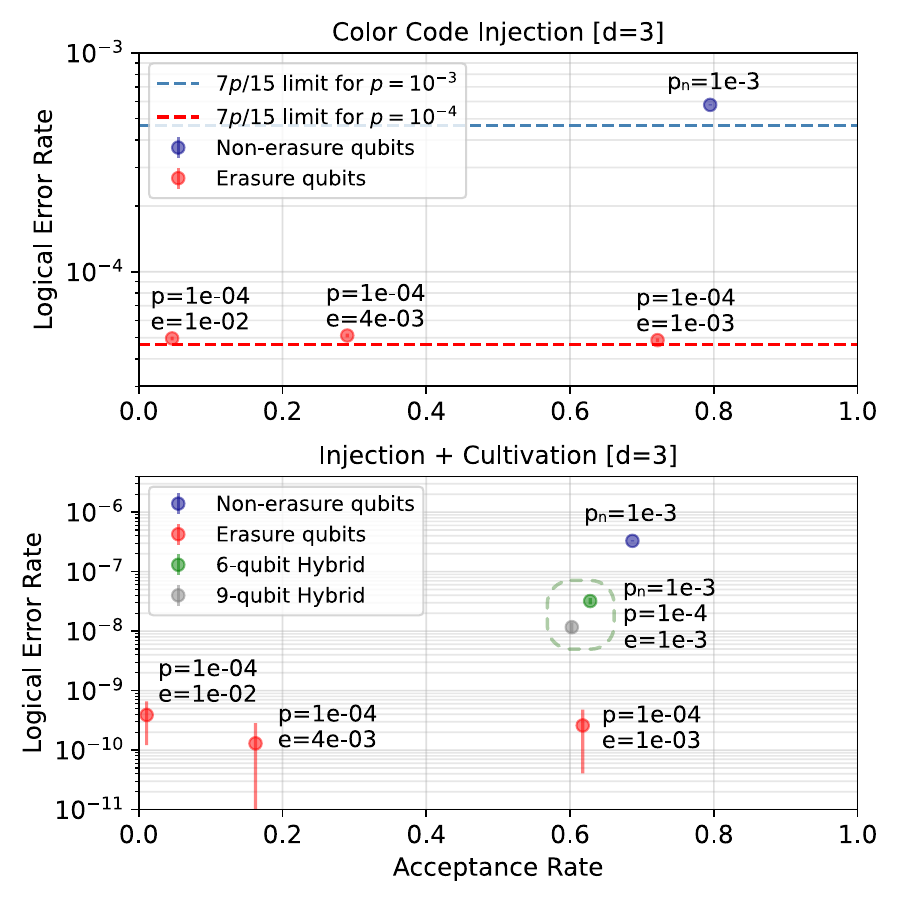}
    \caption{Logical error rate and acceptance rate for unitary injection (a) and unitary injection and cultivation (b) for distance 3 color code, using the protocol form Ref.~\cite{gidney2024magic}.
    Points are for non-erasure ($e=0, p_n=10^{-3}$), almost perfect erasure conversion ($e=10^{-3}, p=10^{-4}$), and two values of non-perfect erasure conversions ($e=4 \cdot 10^{-3}, p=10^{-4}$ and $e=10^{-2}, p=10^{-4}$).
    Hybrid setups with, respectively, 6 erasure qubits and 9 erasure qubits ($e=10^{-3}, p_n=10^{-3}, p=10^{-4}$) are also presented.
    The theoretical limits [Eq. \eqref{eq:logical-error-rate-color}] for non-erasure qubits and erasure qubits are depicted by the dashed blue and red lines (a), respectively.
    }
    \label{fig:injection-cultivation}
\end{figure}

The Magic State Cultivation protocol's primary advantage, however, emerges during the second step, cultivation, where we perform a fault-tolerant measurement of the logical $H_{XY}\equiv (X+Y)/\sqrt{2}$ operator on the color code. This critical measurement detects low-order errors that would otherwise dominate the logical error rate, substantially improving the overall protocol's effectiveness. By identifying and flagging these errors, the fault-detection mechanism significantly enhances the fidelity of the resulting magic state beyond what the initial injection alone can achieve.

The arguments presented in Section \ref{sec:injection-with-erasure}, which lead to the conclusion that the logical fidelity is set by the residual Pauli error rate of the erasure qubits, Eq.~\eqref{eq:erasure-logical-rate}, and that the acceptance rate is roughly equal to that of non-erasure qubits with the same noise strength, Eq.~\eqref{eq:erasure-success-rate},
are applicable to the first two steps, injection and cultivation, of the overall cultivation procedure. Accordingly, we simulate these steps and present the results in Fig. \ref{fig:injection-cultivation}. 
In this section, for better comparison with Ref.~\cite{gidney2024magic}, we adopt the same uniform noise model as used there. In this model, each operation (1Q, 2Q, SPAM) is subject to a Pauli error with probability $p$ and an erasure error with probability $e$. While these simulations do not incorporate erasure measurement errors, our analysis in Section \ref{sec:injection-with-erasure} suggests that failed erasure measurements would likely be mitigated by subsequent erasure and stabilizer measurements in the protocol. We anticipate this limitation would manifest only as a small reduction in overall performance~\cite{Bailey_optimization_2024}.

In the final grafting step, the code is expanded into a larger distance hybrid color and surface code. Here, an abort criterion is determined by gap decoding~\cite{gidney2024magic}. 
As shown in \cite{vaknin2025cultivation}, erasure qubits will give an advantage to this step as well, since erasure information can be used to refine the soft abortion criterion. 
On the other hand, in architectures where erasure qubits are costly compared to non-erasure qubits, it may also be pertinent to consider a hybrid approach where a small color code patch of erasure qubits is grafted into a larger patch of non-erasure qubits.
We leave these more complex questions for future work.

Unlike injection, the cultivation circuit does not have any single fault location capable of undetectably causing a logical error.
Instead, logical errors arise through combinations of faults, each involving at least $d_1$ fault locations.
For $d_1=3$, a computerized search reveals hundreds of thousands of such fault combinations, making it unfeasible to fully suppress errors by converting a small subset of qubits to erasure qubits, as was done for injection in Section~\ref{sec:Hybrid-architecture}.
Nevertheless, we find that converting just 6 qubits into erasure qubits can suppress at least one fault in most of these combinations, resulting in over a tenfold reduction in the logical error rate. 
Further improvement is achieved by converting 9 qubits into erasure qubits, fully covering at least one fault in every fault combination and reducing the error rate by a factor of thirty.
Both hybrid schemes are illustrated in Fig.~\ref{fig:injection-cultivation} as the ``6-qubit Hybrid'' and ``9-qubit Hybrid'' data points.
However, the benefit of these hybrid schemes is modest, as in contrast converting all 15 qubits to erasure qubits yields a significantly greater improvement --- by roughly a factor of $10^3$.

\section{Conclusion}

We have demonstrated that erasure qubits can significantly enhance magic state injection and magic state cultivation, by providing additional information that enables more effective post-selection. 
Specifically, the logical error rate can drastically improve with minimal reduction in the acceptance rate, if dominant errors can be converted into detectable erasure errors.
This improvement persists even when erasure rates are modestly higher than error rates in non-erasure qubits, and erasure detection itself is noisy.

For surface code injection, we moreover demonstrate that almost all the benefits of using erasure qubits can be achieved using a hybrid system with only three strategically placed erasure qubits at the most impactful locations, offering an efficient alternative
when full erasure qubit implementation proves cost-prohibitive or introduces significant performance overhead.

For color code cultivation, on the other hand, our analysis shows that it pays to make all the qubits of the injection and cultivation process into erasure qubits.
In this case, for cultivation on a $d=3$ color code, we find that erasure qubits with erasure rate $e \gtrsim 10^{-3}$ and residual error rate $p=10^{-4}$ have approximately a thousandfold improvement in logical error rate over non-erasure qubits with physical error rate $p_n=10^{-3}$, at a marginal cost in acceptance rate.

The benefits of using erasure qubits become even more pronounced when considering subsequent code expansion, which we study for injection into the surface code. We speculate that large gains can also be had for the ``grafting'' step of magic state cultivation, see also a discussion in Ref.~\cite{vaknin2025cultivation}, but leave a quantitative study of this for future work. Moreover, the method presented here can be directly implemented to cultivation methods that utilize non-local connectivity \cite{vaknin2025cultivation, chen2025efficientmagicstatecultivation}. Our results for the injection and cultivation steps are, however, indicative that extremely low logical error rates relevant for early fault-tolerant applications may be within reach with erasure rates $\sim 10^{-3}$ and residual undetectable errors at the $\sim 10^{-4}$ level.


 \begin{acknowledgments}
We thank the staff from across the AWS Center for
Quantum Computing that enabled this project.
\end{acknowledgments}

\appendix

\section{Noise model} \label{sec:noise-model}

In our simulations, we employ a noise model in which two-qubit gates (CNOT), state-preparation and measurement operations (SPAM) carry a probability $p$ of inducing a Pauli error, and non-SPAM one-qubit operations (idling and single qubit gates) carry a probability $p/10$ of inducing a Pauli error.
Idling occurs at every stabilizer extraction round in parallel with measurements, and affects all qubits not being measured. 
All Pauli errors are considered equally likely, i.e., the noise channels are fully depolarizing.

For erasure qubits, there is an additional probability $e$ of erasure error occurring for two-qubit gates and SPAM operations, and a probability of $e/10$ for erasure during non-SPAM one-qubit operations. 
An erased qubit that interacts with another qubit induces a fully depolarizing channel on the other qubit. A measurement on an erased qubit gives a random outcome, while state preparation is assumed to correctly re-initialize a qubit, even if it was previously erased.
In two-qubit gates, each qubit is erased with probability $1-\sqrt{1-e_2}$, leading to a total erasure rate of $e_2$ per gate.

\section{Injection types} \label{sec:injection-types}

\subsection{Lao-Criger injection}

The protocol in Ref.~\cite{lao2022magic}, which we refer to as ``Lao-Criger injection,'' is a version of the protocol presented by Li in Ref.~\cite{li2015magic},  adapted to the rotated surface code. It starts by initializing a qubit in the desired injection state, and measuring the stabilizers $r$ times.

The initialized qubit can be either in the corner of the patch or in the middle of the patch. In this study, we focus exclusively on initializing the center qubit, as it exhibits a better error rate, see~Ref.~\cite{lao2022magic} for further details.

The logical error rate up to first order in the gates error is:
\begin{align}
    p_{\text{LI}} = \frac{3}{5} p_2 + p_{IN} + \frac{2}{3} p_1 + O\qty(p^2) \;.
\end{align}

This error rate results from the rate of initialization error affecting the middle qubit, a single-qubit gate error during the rotation of this qubit into the magic state, and 9 particular two-qubit errors, each occurring with probability $\frac{p_2}{15}$ in three specific CNOTs during the first round of stabilizer extraction.

The noise channel is:
\begin{multline}
    \mathcal{E}_\text{LI}  = (1-p_\text{LI}) [I] + (p_\text{IN} + \frac{2}{3} p_1) [F_L]
    \\
    + \frac{5}{15} p_2 [X_L] + \frac{2}{15} p_2 [Y_L] + \frac{2}{15} p_2 [Z_L] \;,
\end{multline}
with $[F_L]$ being the channel that flips the state to an orthogonal state, and $[P_L]$ being the channel that applies the logical Pauli $P$ on the state \cite{li2015magic}.

\subsection{Hook injection} \label{sec:hook-inject-appendix}

Hook injection was introduced by Gidney in Ref.~\cite{gidney_hook_injection}. Instead of initializing a qubit in the desired injection state, it initializes a $\ket{+}$ state and rotates the logical operator. It does so by switching the order of the CNOTs in the stabilizer extraction circuit, in a way that $Z$ rotations on an ancilla rotates the logical operator. 

\begin{figure}[t]
    \centering
    \includegraphics[width=\linewidth]{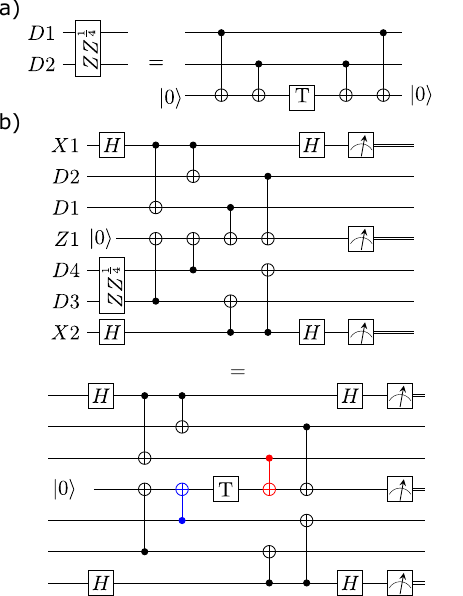}
    \caption{(a) The identity from Eq. \eqref{eq:rotation_identity} for $\theta=\frac{\pi}{8}$. 
    (b) Propagating the $T$ gate using this identity.}
    \label{fig:T-gate-identity}
\end{figure}

This can be seen by propagating the $Z$-rotation operator to the start of the circuit, Fig. \ref{fig:T-gate-identity}.
The propagation itself can be derived by observing the relation:
\begin{multline} \label{eq:rotation_identity}
    \prod_{i=1}^n (\text{CNOT}_{ia}^\dagger) \cdot e^{i \theta Z_a} \cdot \prod_{j=1}^n (\text{CNOT}_{ja}) 
    \\
    = e^{i \frac{\pi}{4} \sum_i(I_i-Z_i)\otimes (I_a-X_a)}
    \cdot e^{i \theta Z_a}
    \cdot e^{i \frac{\pi}{4} \sum_i(I_i-Z_i) \otimes (I_a-X_a)}
    \\
    = e^{i\theta Z_1 \dots Z_n Z_a} \;,
\end{multline}
for a set of qubits indexed from $\{1,\dots,n\}$ and ancilla indexed as $a$. 
If the ancilla is initialized in $\ket{0}$, we get the relation depicted in Fig. \ref{fig:T-gate-identity}a.

The hook injection has a better logical error rate, contributed only from one error mechanism in the $T$ gate application, and three error mechanisms in the following and preceding $\text{CNOT}$ gates (blue and red in Fig. \ref{fig:T-gate-identity}b). This gives an error rate of
\begin{align}
    p_\text{HI} = \frac{1}{3}p_1 + \frac{3}{15} p_2 \;.
\end{align}

The noise channel is
\begin{multline}
    \mathcal{E}_\text{HI}  = (1-p_\text{HI}) [I] + (p_\text{IN} + \frac{1}{3} p_1) [F_L]
    \\
    + \frac{2}{15} p_2 [Y_L] + \frac{1}{15} p_2 [Z_L] \;.
\end{multline}

\section{Simulating erasure qubits} \label{sec:simulating-erasure-qubits}

In this paper, we use the Stim library \cite{gidney2021stim} to perform Clifford simulations. In order to simulate erasures we generate an erasure sequence for each run, and insert fully depolarizing noise wherever an erasure has happened, converting the erasure circuit into a matchable stabilizer circuit. After which, we perform minimum-weight perfect matching decoding using PyMatching \cite{higgott2022pymatching}.

Throughout the paper, we inject an $\ket{S}=S\ket{+}=(\ket{0} + i\ket{1})/\sqrt{2}$ state as a test case for injection, as it includes only gates that can be simulated efficiently on classical computer \cite{knill1998resilient}.

In section \ref{sec:injection-with-erasure}, we perfectly decode the resulting injected state by performing a noiseless stabilizer extraction round at the end of the circuit.

In Section \ref{sec: expanding-the-state}, we employ the circuits published by Gidney in Ref.~\cite{gidney_hook_injection}, which consist of $r$ rounds of injection, a single round of code expansion, and $d_2$ memory rounds, followed by a in-place $Y$-basis measurement \cite{Gidney_2024inpace-Y} to verify the injected state. In the sole expansion simulation (Figure \ref{fig:expansion-and-end2end}a), we sample the circuits with noise applied only during the expansion step. In the end-to-end simulations (Figure \ref{fig:expansion-and-end2end}b), noise is applied across all operations throughout the circuit.

Throughout the paper, error bars represent 95 percent confidence intervals.

\begin{figure}
    \centering
    \includegraphics[width=\linewidth]{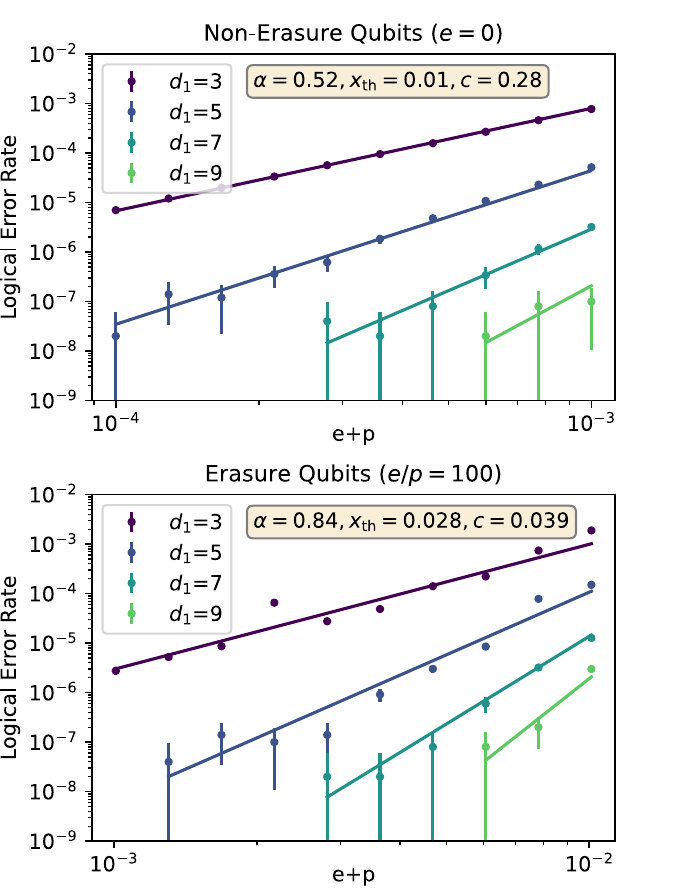}
    \caption{\label{fig:expansion-fit-appndx}Logical error rate of the noisy expansion step from a distance-$d_1$ code to a distance-$d_2$ code, with the low-distance code state being prepared noiselessly. Solid lines represent fits using Equations~\eqref{eq:ansatz-erasure-appndx} and \eqref{eq:ansatz-non-erasure-appndx}, with parameters given in the floating text box.
}
    \label{fig:enter-label}
\end{figure}

To investigate the scaling with distance for the expansion step, c.f. Eq.~\eqref{eq:ansatz-erasure}, we fix the ratio of erasure to Pauli errors, such that $p=Re$ for a constant $R$, and define $x= e + p = (1 + R)e$ to have a one-parameter noise model.
For small $x$ we use an ansatz from Ref.~\cite{Bailey_optimization_2024} for the logical error rate
\begin{align} \label{eq:ansatz-erasure-appndx}
    p_{L,\text{exp}} = c \qty(\frac{x}{x_\text{th}})^{\alpha d} \;,
\end{align}
with $c$, $x_\text{th}$, and $\alpha$ fit parameters.
For non-erasure qubits, a slightly better fit is achieved with
\begin{align} \label{eq:ansatz-non-erasure-appndx}
    p_{L,\text{exp}} = c \qty(\frac{x}{x_\text{th}})^{\alpha (d+1)} \;.
\end{align}
Note that for the code expansion, we are limited by the smallest distance $d_1$ in the first round of the expansion, and we thus expect the logical error rate to be approximated by these ansätze for $d=d_1$.

In Fig. \ref{fig:expansion-fit-appndx}, we fit those ansätze to numerical simulations of expansion where the low-distance code $d_1$ is prepared noiselessly. 

\section{Acceptance rate} \label{sec: success-rate}

The probability of not having any erasure, is given by
\begin{align}
    \text{AR}_e (\bar{e}) = (1-e_\text{SPAM})^{x_\text{SPAM}} (1-e_1)^{x_1} (1-e_2)^{x_2} \;,
\end{align}
where $x_1$ ($x_2, x_\text{SPAM}$) is the number of 1Q (2Q, SPAM) gates in the circuit.
The probability of not having any detected Pauli error is more complex, as for each gate some Pauli combinations will be detected and some will not.
Denoting the probability of a detected Pauli error on gate $g$ to be $p^D_g<p_g$ with $p_g$ the error rate of the gate, the acceptance rate can be approximated:
\begin{align}
    \text{AR}_p (\bar{p}) \simeq \prod_{g\in{G}} (1-p^D_g) \;.
\end{align}
with $G$ being the set of all gates in the circuit.
This expression approximates the acceptance rate by considering only first-order error probabilities, presuming that higher-order terms
have a negligible impact on the overall probability.

\begin{figure}
    \centering
    \includegraphics[width=\linewidth]{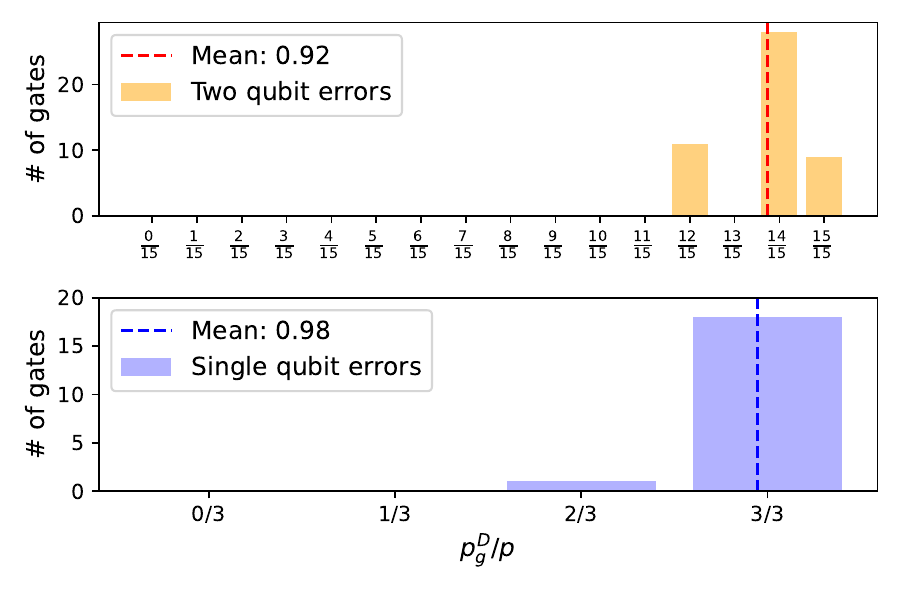}
    \caption{Histogram of detection rate of errors in two-qubit gates (top) and single qubit gates (bottom) for distance 3 hook injection circuit with two stabilizer extraction round $r=2$. SPAM errors are all detected and thus not displayed in the figure.
    }
    \label{fig:error_mechanisms_hist}
\end{figure}

To evaluate the probability for a Pauli error to be detected, we show a numerical search of the probability for detection per gate, $p_g^D$ for the $d_1=3$ hook injection circuit in Figure \ref{fig:error_mechanisms_hist}. Each error location is analyzed for the number of error mechanisms (i.e the number of different Pauli combinations that are detected) and the normalized probability for each gate is counted in the histogram. 
It can be seen that for both 2Q gates and 1Q gates, most error mechanisms are detected. SPAM errors are all detected and thus not displayed in the figure.

\section{Hybrid dual-rail and transmon architecture} \label{sec:Hybrid-dual-rail-and-transmon-architecture}

We here motivate an error model for the hybrid non-erasure and erasure qubit protocol from Sec.~\ref{sec:Hybrid-architecture}. We base this on a scenario where the non-erasure qubits are transmons and the erasure qubits are dual-rail qubits.
Since dual-rail qubits and transmons have different gate times, to implement the hybrid architecture, we need to account for idling of some qubits while waiting for other gate to finish. In this section, we offer a general analysis of the idling errors for the hybrid architecture.

In the hybrid patch, we have 3 types of two-qubit gates, (1) between two transmons, (2) between transmon and dual-rail (3) between two dual-rails. Each of those has different gate time: $t_\text{2dr}$ between two dual rails, 
$t_\text{2tr}$ between two transmons, $t_\text{tr-dr}$ between a transmon and a dual rail. Dual-rail gates have longer gate times \cite{Kubica_2023}, thus in each step of the stabilizer extraction circuit, transmons participating in a transmon-transmon gate need to idle while waiting for the dual-rail gates to finish.

Our assumptions are:
\begin{itemize}
    \item Since the dual-rail and transmons are fabricated on the same chip, we assume they share the same decay time $T_1$.
    \item We assume equal dephasing and decay times $T_\phi=T_1$ for the transmon and neglect dephasing for the dual-rail.
    \item The error rates of those gates are dictated only by the decay and dephasing rates of the qubits. This assumption is not exact, however, due to the balance between gate fidelity and gate time, it is often appropriate \cite{Motzoi_2009}.
\end{itemize}

Under the twirling approximation, the Pauli error rate for transmons is \cite{Ghosh_2012}:
\begin{align}
    p(t) = 3/4 - \frac{1}{4}e^{-t/T_1} - \frac{1}{2}e^{-3t/2T_1} \approx t/T_1 \;,
\end{align}

The erasure rate of dual-rail qubits can be expressed by 
the decay rate of the two transmons it is composed of.
Since dual-rail encoding maintains exactly one excitation distributed between the two transmons \cite{Kubica_2023}, the erasure rate follows:
\begin{align}
    e(t)
     = 1 - e^{-t/T_1} \approx t/T_1 \;.
\end{align}

During idle periods, it can be assumed that the transmons undergo dynamic decoupling, reducing most dephasing impact.
Thus, when a transmon remains idle for a duration $t_\text{idle}$, it experiences an error rate of
\begin{align}
    p_\text{idle} (t_\text{idle}) = t / 2T_1 \;.
\end{align}

In each timestep of the stabilizer extraction circuit, the transmons are subject to two distinct error sources: the two-qubit gate operations ($p_{n,2}$) and the subsequent idling period ($p_{n,\text{idle}}$) while waiting for the dual-rail operations to complete. The combined effective error rate $p^*_{n,2}$ can be expressed as:
\begin{multline}
    p^*_{n,2} = p_{n,2} + p_{n,\text{idle}} \\ = p(t_{2\text{dr}}) + \frac{t_{2\text{dr}}- t_{2\text{tr}}}{2 T_1} = \frac{e_2 + p_{n,2}}{2}.
\end{multline}

SPAM operation can be delayed or advanced to address the different gate time issue, and one-qubit gates are assumed to require negligible idling time as they are shorter with smaller error rates.


\providecommand{\noopsort}[1]{}\providecommand{\singleletter}[1]{#1}

\end{document}